\documentclass[conference]{IEEEtran}
\usepackage{cite}
\usepackage{amsmath,amssymb,amsfonts}
\usepackage{algorithmic}
\usepackage{graphicx}
\usepackage{textcomp}
\usepackage{xcolor}
\def\BibTeX{{\rm B\kern-.05em{\sc i\kern-.025em b}\kern-.08em
    T\kern-.1667em\lower.7ex\hbox{E}\kern-.125emX}}
\begin{document}

\title{Expanding ML-Documentation Standards For Better Security
}

\author{\IEEEauthorblockN{1\textsuperscript{st} Cara Ellen Appel}
\IEEEauthorblockA{
\textit{Universität Hamburg}\\
Hamburg, Germany\\
ORCID 0009-0006-8113-428X}

}

\maketitle

\begin{abstract}
 This article presents the current state of ML-security and of the documentation of ML-based systems, models and datasets in research and practice based on an extensive review of the existing literature. It shows a generally low awareness of security aspects among ML-practitioners and organizations and an often unstandardized approach to documentation, leading to overall low quality of ML-documentation. Existing standards are not regularly adopted in practice and IT-security aspects are often not included in documentation. \par 
 Due to these factors, there is a clear need for improved security documentation in ML, as one step towards addressing the existing gaps in ML-security. To achieve this, we propose expanding existing documentation standards for ML-documentation to include a security section with specific security relevant information. Implementing this, a novel expanded method of documenting security requirements in ML-documentation is presented, based on the existing Model Cards and Datasheets for Datasets standards, but with the recommendation to adopt these findings in all ML-documentation. \par  
\end{abstract}


\section{Introduction}
Every day, millions of cyberattacks are launched against internet users and organizations \cite{Microsoft45}. Cyberattacks aim to disrupt or take over control of systems and networks, as well as to steal, exploit or delete data involved in them \cite{Vakil26}.
The field of IT-security aims to protect the confidentiality, integrity and availability of data and systems \cite{Tsochev25}\cite{Khan51}. To achieve this, a number of different protections need to be implemented to secure any IT-system \cite{Vakil26}\cite{Jevtic24}. \par
AI has only become well-established in everyday technology usage in the past few years, as publicly accessible AI-technologies have become available directly to consumers. But the idea of creating artificially intelligent systems is an old one \cite{shao30}.  What largely contributed to the widespread application of AI within the last decade is the advent of machine learning, which significantly accelerated AI development and progress \cite{shao30}. The importance and societal impact of  AI continues to grow with it being increasingly used in critical systems \cite{Boenisch10}\cite{Wang43}. A failure in such a system might result in catastrophic consequences like the loss of human lives \cite{Wang43}. Machine learning is currently at the core of most artificial intelligence systems, and is one of the most developed and widely applicable fields in AI-research \cite{Humm42}. Organizations rely on ML-models more than ever for their operations, which has led to an ever-greater need for robust security to protect ML-systems \cite{Khan23}.\par Documentation is one area where steps towards better ML-security can be made. Security as a critical aspect of software quality should be discussed in documentation, although in practice this often is not the case, both for ML and traditional software \cite{Khan51}. But existing ML-documentation standards do not include security aspects and these standards often are not used by practitioners \cite{Mitchell9}\cite{gebru41}. Practitioners similarly do not include security aspects in ML-documentation in practice, due in large part to their lack of security awareness regarding ML-systems \cite{Boenisch10}\cite{Grosse11}. We address the resulting clear need to close the gaps in ML-documentation regarding security in this paper, as well as discussing the larger problems facing ML-security.

\section{Methodology}
We have conducted an extensive review of the existing literature in three main areas: ML-documentation (both its implementation in practice and the existing standards), ML-security and traditional software documentation. To identify relevant literature, we used Google Scholar due to its broad coverage of recent academic literature and its keyword search functionality. We searched for literature that was published from 2019 onwards, as a preliminary search of commonly cited research in these fields showed that almost all relevant research was conducted since then. We searched for articles with the following keywords, as we identified these as relevant to the topic of this paper: ‘ML’, ‘AI’, ‘standards’, ‘security’, ‘Adversarial Machine Learning’, ‘model’, ‘dataset’, ‘software documentation’. We searched for these terms individually as well as in various combinations like ‘ML security documentation’. Furthermore, we looked at the cited research in the articles we found through our search and, when relevant, also included these in our literature review.\par
Chapters III to VI of this article discuss the state of the research based on our literature review. Chapter VII presents our proposed method of expanding existing ML-documentation standards to include security aspects, with the goal of raising the quality of security documentation for ML. This section draws on the results of the previous chapters. In chapter VIII we discuss challenges in implementing the proposed method, before concluding with recommendations for future work in chapter IX.

\section{The Role of Documentation in IT-Security}
There are many different documentation standards for non-ML software, serving different purposes. These help practitioners create suitable high-quality documentation \cite{Konigstorfer6}. But while documentation guidelines for ML exist, they do not provide adequate guidance for practitioners on the same level as guidelines for traditional software do \cite{Konigstorfer6}. 
One of documentation's main goals is for its corresponding software to fulfill all legal, compliance and technical requirements \cite{Konigstorfer6}. Documentation is the main resource used by practitioners to understand and evaluate software \cite{Bhat12}. State of the art in documentation is the automated generation of software documentation with only minimal effort required from developers \cite{Hashemi4}. But this is not yet the reality for ML-documentation, even though some researchers have proposed more interactive methods of documentation for ML-models \cite{Crisan14}. \par
Documentation covers different aspects of a software system, including security. In regards to documenting the security aspects of a software system, one option is the creation of specific security documentation with a great level of detail regarding IT-security, the other is including security aspects in non security-specific documentation. This can help encourage all developers to take security considerations into account and reliably document their projects with security in mind \cite{Gorski31}. Thus, a combined approach of using and creating both specific security documentation and more general documentation, which includes security concerns on a more basic level, is optimal \cite{Gorski31}. \par 
The reality for ML-documentation is however a distinct lack of specific security-documentation standards, as well as the inclusion of few security aspects in general ML-documentation. This is the case both within existing standards, as well as within documentation created by practitioners \cite{Gupta19}.

\section{The Current State of Security for AI and ML}
Both AI technologies and the world they exist in and draw data from are continuously changing and evolving, which makes safeguarding AI and more specifically ML a challenging task \cite{Leslie1}. Cyberattacks on ML-systems occur significantly more frequently than many practitioners are aware of\cite{Microsoft55}. But, security continues being a minor factor in ML-development, often overlooked in favor of functionality, which is generally seen as most important \cite{Boenisch10}. Because of inadequate standards for securing ML-models and -systems as well as due to a lack of awareness from organizations, the security of ML-systems still depends largely on the awareness and education of the involved practitioners, which studies show is generally quite low \cite{Boenisch10}\cite{Grosse11}. \par
Within IT-security, both attackers and defenders operate with a balance of costs and benefits that their attacks or security measures create in mind \cite{Grosse15}. As Apruzzese et al. state, economics are the main driver of IT-security, which if overlooked may lead to the creation of unrealistic solutions \cite{Apruzzese20}. Every security mechanism aims to increase the cost of an attack for the attackers as much as possible, for the lowest possible cost for the defenders \cite{Apruzzese20}.
There is a significant disconnect between the settings and threats studied in ML-security research and what practitioners experience in the real world, as multiple research groups have observed \cite{Grosse15}\cite{Bieringer16}\cite{Apruzzese20}. They have identified a number of areas where reality diverges from what is currently studied, among them the following: 
\begin{itemize}
    \item Research focuses on securing models, whereas models are parts of larger ML-systems and pipelines in practice \cite{Grosse15}\cite{Bieringer16}\cite{Apruzzese20}.
    \item Research on the impact of attacks look only at specific settings, which do not align with real-world applications and attacks \cite{Grosse15}.
    \item The majority of research into attacks focuses on evasion attacks, whereas practitioners are most concerned about poisoning attacks and model theft \cite{Bieringer16}.
    \item Not all vulnerabilities occur in development, the phase most often studied regarding ML-security. There is a lack of research covering the vulnerabilities faced once models and systems are deployed \cite{Gupta19}.   
\end{itemize}
Additionally, researchers' results and recommendations often do not reach practitioners. Multiple studies have identified generally low rates of awareness from practitioners regarding the need for ML-security and how to achieve it \cite{Boenisch10} \cite{Grosse11}. In the fittingly-named paper “‘I never thought about securing my Machine Learning systems’ - A study of security and privacy awareness of machine learning practitioners”, Boenisch et al. found that while almost two thirds of the practitioners they interviewed rated security as important or very important, 24\% of them had not heard of any of the mentioned common types of attacks (inversion, evasion, impersonation and poisoning) \cite{Boenisch10}. Even practitioners that are aware of the risks that stem from threats and attacks struggle to evaluate the security of their own models and systems \cite{Nahar13}. A significant reason for this is how little training on ML-security universities still provide to their students who then become practitioners \cite{Boenisch10}. Researchers also identified lacking differentiation between ML-security and general IT-security, which can result in overestimating established security strategies \cite{Bieringer16}. \par 

\subsection{Threats to ML-Security}
ML-systems face many different types of threats and attacks, which can vary in severity and purpose. The level of risk a system faces varies based on a number of factors, which can be assessed through threat models. But regardless of the individual risks for an ML-system, security considerations should be kept in mind at every stage of development. \par
Attacks against ML are either exploratory or exploitative in nature. Exploratory attacks are passive attacks, where no disruption to the model or system is intended, and the motivation is to extract information from the data or model. They mainly aim to recreate the functionality of a model or system \cite{Koball22}. Exploitative attacks on the other hand are active attacks where disruptions are intended through the manipulation of training or inference data and other means. They mainly aim to disrupt the system to make it unusable \cite{Koball22}. Both exploratory and exploitative attacks can be either targeted or untargeted attacks. Targeted attacks utilize adversarial examples that have been created to make an ML-system do a specific action, whereas untargeted attacks aim to cause general harm to the system and do not have a specific goal \cite{Malik21}. \par
Threat models are used to identify the relationships between attackers and their targets for security assessments. An attacker's behavior is characterized based on their knowledge, capabilities and goals \cite{Apruzzese20}. An attacker's access to knowledge of an ML-model's datasets, features and algorithms can vary between extremely limited knowledge (black-box attacks), full access to information in the case of an internal attacker (white-box attacks) and limited knowledge in some areas and more in others (gray-box attacks) \cite{Koball22}. The capabilities of an attacker describe the types and quantities of data and features in a model that they can alter \cite{Grosse15}. Lastly an attacker's goals are defined by the area or areas of security that they intend to harm, which can be availability, integrity or confidentiality \cite{Grosse15}. These three aspects depend on both the attacker's economic situation, relationship with the attacked organization, country of origin and more, as well as the organization's security protocols, size, type and amount of data they work with, as well as other factors \cite{Apruzzese20}. \par 
There are six most common types of attacks against ML-systems, which are poisoning, backdoors, evasion, model stealing, membership inference and attribute inference \cite{Grosse15}. Recently, there has also been a rising number of sponge attacks \cite{Bieringer16}. Poisoning, backdoor, evasion and sponge attacks are exploitative in nature, as they target availability and accuracy, whereas model stealing and inference attacks are exploratory. Depending on their level of access, capabilities and goals, attackers will choose between these different attack types.

\subsection{Securing ML}
The goals and objectives in ML-security largely align with those of traditional IT-security \cite{Grosse15}. However, the strategies and measures used to achieve these goals differ greatly between traditional IT-systems and those based on ML. As data forms the core of all ML-based technologies, securing data collection, storage and preparation processes are additional objectives when securing ML \cite{Zhang2}. And due to the less static nature of ML-systems compared to traditional IT-systems, robustness is another important objective \cite{Leslie1}. \par
Security measures for ML-systems depend on many factors. Among them are the scale and scope of the system, the types of models involved, the amount of (especially sensitive) data they use, as well as the sources of that data \cite{Leslie1}. There are multiple possible countermeasures that an organization can use to protect their models and systems against each type of attack. Deciding which of these to utilize depends not just on the above mentioned factors, but also on decisions an organization needs to make regarding the trade-offs that any countermeasure entails \cite{Malik21}. These trade-offs are most commonly between security and cost, and between accuracy and robustness. Employing these countermeasures always creates additional costs, and often leads to a less accurate but more robust model or system \cite{Malik21}. Thus, the decision of which mitigations to employ is always in part an economical one \cite{Apruzzese20}.\par
Countermeasures may protect against attacks through specific vectors like the input data, but can also work to increase the overall security of a system, either with a specific goal like safeguarding privacy or by increasing security as a whole. Some of the possible efforts to protect security in general are differential privacy, security auditing and techniques like PATE, the Privacy Aggregation of Teacher Ensembles which allows models to be trained on sensitive data, without that data being linked to individuals \cite{Malik21}. \par
There are also frameworks for ML-security created by organizations like the MITRE corporation or the US National Institute of Standards and Technology. Such frameworks can be a valuable resource to enhance ML-system security as they provide a structured approach to risk management and mitigation, and offer best practices going over the entire development lifecycle \cite{Malik21}. Both the MITRE and NIST frameworks offer practical guidance to increase security in ML-projects. However, their use requires a proactive approach to security from organizations and practitioners, as well as a high level of training, which is often not present in practice \cite{Malik21}. 

\section{Documentation for Traditional Software}
Documentation needs to be kept up-to-date and be continuously worked on to accomplish its purpose and be a useful tool \cite{kruse48}. This is hindered by documentation often being seen as a low priority by developers, which is why automated approaches which use AI to create documentation are being developed\cite{Hashemi4}\cite{kruse48}. Despite the important role documentation plays for software, there are common shortcomings in documentation created by practitioners \cite{Aghajani47}. A 2020 survey of 146 practitioners by Aghajani et al. shows that documentation commonly contains errors and missing information with a significant impact on documentation quality and usefulness. The main reason given in the paper was a lack of time for creating documentation \cite{Aghajani47}. \par
Security is a critical aspect of software quality, and therefore an important area to include in software documentation and should be considered in every step of the development process \cite{Khan51}. This can be achieved by addressing security through Requirements Engineering (RE), which is the process of specifying, systematically analyzing and fine-tuning requirements that a software system needs to fulfill \cite{Hofmann53}. RE can be a part of the software documentation process, as the resulting requirements allow stakeholders to easily identify and understand the software's purpose and goals, which is one of the aims of documentation \cite{Hofmann53}. However, Khan et al. have identified an ongoing lack of consideration for security aspects in RE in their 2021 systematic literature review. They found that security requirements were commonly not properly negotiated (46\% frequency) and not validated (40\% frequency), as well as lacking risk assessment and analysis (28\% frequency) and developing threat modeling (21\% frequency) \cite{Khan51}. \par
One of the reasons for security often not being addressed in software documentation is a lack of concern regarding security risks, leading to security not being a high priority \cite{Khan51}. Security is often only addressed once software development is completed, which leads to a number of risks staying unidentified and unaddressed \cite{Khan51}. Requirements Engineering is also often not utilized in software projects, as it is hard-to-budget-for and there are not many dedicated RE professionals involved in software projects \cite{maalej52}. As RE is one of the documentation processes in which security is most often addressed, this further complicates security documentation for software systems. Further research into Requirements Engineering for ML and how to encourage practitioners and organizations to adopt it in practice is needed, so that RE can help address the identified gaps in ML-security and documentation.

\section{Documentation for ML}

Documentation standards that are used for traditional software do not work well for ML-documentation, due to the many differences to traditional software. Some of the differences with the largest impact on documentation are the limited meaning of the source code in ML, its black-box nature, the high dependency on data and frequent changes that result from this, as well as the higher human influence on ML-projects \cite{Konigstorfer6}. But many of the issues seen in traditional software documentation are also present in documentation for ML-systems, and the reasons for why documentation is needed are also largely the same \cite{Liang7}.\par
Thorough and effective model documentation is essential for successful communication between developers and users, as documentation can give guidance on potential biases, errors, limitations and the general functionality of an ML-model \cite{Liang7}. Additionally, it can aid in preventing common problems with ML-projects, like not repeating mistakes, avoiding technical debt, preserving institutional knowledge and enabling reproducibility \cite{Chang8}. Implementing industry wide documentation standards for ML can help mitigate the harm that these technologies can and do cause and ensures responsible and ethical deployment \cite{Liang7}. The potential for harm is especially significant when impacted people are unable to contest and interrogate decisions made by ML-systems. This kind of situation can be prevented by creating good and thorough documentation that allows for a better understanding of ML-systems \cite{Crisan14}. \par
While documentation standards and frameworks for ML-projects have been proposed and model hosting sites and companies encourage users to document their models with them, the use of these standards is still limited and documentation created with them often incomplete \cite{Liang7} \cite{Bhat12}. Some of the most widely known proposed documentation standards are ML Model Cards (proposed by Mitchell et al. in 2019), FactSheets (proposed by Arnold et al. in 2018), Datasheets for Datasets (proposed by Gebru et al. in 2021) and Dataset Nutrition Labels (proposed by Holland et al. in 2018) \cite{Gupta19}. Model Cards and FactSheets focus on model documentation, whereas Datasheets for Datasets and Dataset Nutrition Labels focus on documenting datasets. \par
Requirements Engineering can also be adapted for ML-systems, but the already present lack of RE professionals in traditional software projects is even more significant in ML-projects \cite{maalej52}. Maalej et al. addressed Requirements Engineering for AI systems in a 2023 paper, specifically focusing on using it to create Responsible AI \cite{maalej52}. Security, both technical systems safety and data privacy, is included in the seven core requirements for responsible AI as proposed by the European Commission's High-Level Expert Group on Artificial Intelligence (HLEG) \cite{EuropeanCommission54}. Maalej et al.'s paper discuss six aspects needing particular consideration for adapting RE for AI projects \cite{maalej52}. Of these, tradeoff analysis, expanding RE to focus on data and requirements as the foundation for quality and testing are particularly relevant for security and need to be addressed in further research, so that RE can be an effective piece of the documentation puzzle for ML-systems. \par 
A 2022 study by Liang et al. into the quality of ML-documentation, specifically when using model cards is encouraged, showed that less than half of the models studied had a model card and the completeness and depth of information varied greatly even when one was available \cite{Liang7}. There were large differences in detail between different sections of a model card, with the training and citation sections filled out most often, while risks, limitations and biases and the environmental impact were only filled out in less than 20\% of all model cards \cite{Liang7}. \par
But why does adoption of these standards lag behind? One important reason mentioned in literature is the amount of effort that creating documentation using these standards requires. This is particularly challenging for non-ML-expert practitioners who due to the widespread inclusion of ML as part of IT-systems are commonly responsible for creating documentation for it \cite{Crisan14}. To aid developers, especially non-experts, in creating good documentation, interactive solutions analogous to those that are already common in non-ML documentation have been proposed which make standards more easily adoptable \cite{Bhat12}\cite{Crisan14}. Interactive approaches also support a deeper interrogation of the model's characteristics in the process of documentation, without adding to practitioner's workloads \cite{Crisan14}. Interactive approaches are preferable to fully automated ones, as automated approaches often introduce errors and biases to the documentation they produce, which reduces the quality of the resulting documentation \cite{Bhat12}. But automated approaches are useful in certain contexts. Having documentation, even of lower quality, is preferable to not having any documentation, so automated approaches can provide a minimum level of documentation for models and datasets where efforts to provide it have not been made. And in research contexts, where scientific papers are often used as a form of documentation, automated approaches can help translate these into more standardized documentation formats like model cards \cite{Singh17}. \par
Current documentation standards for ML-models, datasets and systems as a whole like Model Cards, Data Sheets and others do not include a specific security section or questions that specifically aim to document security-relevant aspects \cite{Mitchell9}\cite{gebru41}. While aspects included in other sections of a Model Card or Data Sheet like data sources, model details or use cases are relevant in security considerations, they do not sufficiently document all relevant factors. In 2020, Gupta and Galinkin proposed the integration of ethical and security aspects into an automated documentation tool like MLFlow in their paper “Green Lighting ML: Confidentiality, Integrity, and Availability of Machine Learning Systems in Deployment”. They pointed towards the underprepared nature of traditional IT-security for protecting ML-systems and the fact that many vulnerabilities occur during deployment instead of development as reasons for why a greater focus on security in documentation is needed \cite{Gupta19}. This is one example of how security aspects can be addressed in general documentation.  

\section{Proposing New Standards for ML-Security}
As previously discussed, introducing security aspects into ML-documentation is sorely needed. ML-security needs to be improved within the coming years as ML becomes even more ubiquitous and important than it is currently. This is especially important, as awareness for Adversarial Machine Learning and security threats to ML-systems among practitioners is generally still low. Security aspects are still not regularly included in ML-documentation, both in practice and in existing standards. But to improve security awareness and anchor security in ML-workflows, security should be a regular part of documentation for ML-projects. \par
To achieve this, we propose the addition of a security section into all existing ML-documentation standards. The documentation standards of Model Cards, proposed by Mitchell et al., as well as Datasheets for Datasets, proposed by Gebru et al., will be used as an example. We focus on these two standards, as they are the most well-known standards for model documentation (Model Cards) and dataset documentation (Datasheets) \cite{Gupta19}. We will outline what a security section for these standards should look like and how this section would be added to the existing standards.

\subsection{Outlining the Existing Standard for Model Cards}
Model Cards were introduced by Mitchell et al. in 2019 in a paper published by the Association for Computing Machinery out of a need for standardized documentation for ML and have become a widely used standard for ML-documentation as previously described \cite{Mitchell9}. They include a total of nine sections of varying length and detail. These are the following \cite{Mitchell9}:
\begin{itemize}
    \item Model Details (including the model's authors, development date, model version and differences to previous versions, model architecture and type, training information like algorithms, parameters and features, paper or other additional resources, citation details, license information when applicable and contact details of authors)
    \item Intended Use (what are the primary intended uses and who are the primary-intended users, what are out-of-scope use cases and maybe recommendation for other similar models)
    \item Factors (relevant and evaluation factors like technical attributes, environmental conditions or information regarding demographic and phenotypic groups as needed depending on use-cases)
    \item Metrics (like model performance metrics, decision thresholds when used and variation approaches)
    \item Evaluation Data (information about datasets used and why they were chosen and preprocessing of the evaluation data)
    \item Training Data (same information as for evaluation data, if this is possible)
    \item Quantitative Analyses (unitary and intersectional results as disaggregated results)
    \item Ethical Considerations (information regarding possible risks and harms and their mitigations, impact on humans, possible sensitive data or specifically challenging use cases)
    \item Caveats and Recommendations (any additional concerns not covered by the other sections)
\end{itemize}
In these sections security is only included as part of the ethical considerations section in the risks, harms and mitigations, but the focus there is not on security directly. Security might also be mentioned in some cards regarding evaluation and test data sources, although this is not explicitly stated in the Model Cards standard. 

\subsection{Adding a Security Section to Model Cards}
As a tenth section of a Model Card, We propose adding a security section. This section should include information that pertains to the following aspects, as identified in our literature review:
\begin{itemize}
    \item Risk Analysis: Outlines model attributes pertaining to the likelihood and danger of adversarial attacks, including sensitivity of the involved data, how widely it will be deployed, the number of people that will have access to the model and which levels of access they will have, financial and other incentives for attackers like whether the model is used to achieve monetary gain and more.  
    \item Training and Evaluation Data: Includes information regarding the sources of training and evaluation data and whether mitigations like training data sanitization or adversarial training have been implemented.
    \item Model Security: Includes information regarding security mitigations like model pruning, ensemble machine learning or model inspection and sanitization, specifically whether they have been applied to the model, and to what degree and using which methods. This section also includes information on whether the model has been watermarked for model authentification.
    \item Stealing and Inference Attack Mitigations: This section includes information regarding the allowed number of queries to the model in a certain time-frame, whether user behavior-analysis is implemented as well as whether more tradiotional IT-security protections are in place like user-authentication or secure data storage.
    \item Security Testing: Outlines whether security testing of the model like penetration testing has taken place and the results. This section may link to an open-source testing toolkit for ML like Microsoft Counterfit to aid practitioners in security testing their models.
\end{itemize}

\subsection{Outlining the Existing Standard for Datasheets for Datasets}
Datasheets for Datasets were proposed by Gebru et al. as a way to encourage practitioners to document their datasets in 2021 \cite{gebru41}. They are mainly aimed at dataset creators and consumers, but can aid all stakeholders including policy makers, consumer advocates or those whose data is part of the dataset. In the standard outlined by Gebru et al. the following sections and questions make up the datasheet \cite{gebru41}:
\begin{itemize}
    \item Motivation: What purpose was the dataset made for? Who created the dataset and on whose behalf? How and by whom was its creation funded and optional other comments. 
    \item Composition: What are the types of instances in the dataset? Does it contain all possible instances or a sample of a larger set? Is the data raw data or processed? Are labels and targets associated with each instance? Are any instances missing information? Are relationships between individual instances made clear? Are data splits like between training and validation data provided and recommended? Are there any errors, noise or redundancies in the dataset? Is the dataset self-contained or does it rely on external resources via links? Is data that might be seen as confidential part of the dataset like medical data? Does the dataset contain data that might be viewed as offensive, insulting or threatening? \par 
    And for datasets that relate to people also the following questions: Does the dataset identify subpopulations, for example by gender or race? Can individuals be identified through the data in the dataset directly or indirectly? Is sensitive data, like data on religious beliefs, sexual orientation etc. part of the dataset? And any additional comments regarding data relating to people in the dataset.
    \item Collection Process: How was the data acquired? Which mechanisms or procedures were used in collecting the data and how were they validated? What was the sampling strategy if the data set is a subset of a larger set? Which personnel was involved in data collection and how was that personnel compensated? Over what timeframe was the data collected? Were there ethical review processes? \par
    The following questions only apply to datasets that include data relating to people: Was the data related to individual people obtained directly from them or via third parties? Where are the people involved made aware of the data collection? Did they consent to it and the use of their data? Is there a mechanism in place to revoke consent in the future? Has an impact analysis of the potential impact of the dataset been conducted? And any further comments relating to the collection process.
    \item Preprocessing/Cleaning/Labeling: Did the data and the dataset get preprocessed, cleaned or labeled and if so, how was this done? If the data was processed in any way, is the raw data also available? Is the software involved in the processing of the data available? And any further comments on data processing.
    \item Uses: Was the data set used for any tasks already? Do all papers or systems using the data set get collected in a repository that links to them? Which other tasks could the data set be used for? Does the composition of the dataset or its processing impact future uses? Should the dataset not be used for specific tasks? And any other comments on uses of the dataset.
    \item Distribution: Will the data set be distributed to organizations outside of the one for which it was created? How and when will the dataset be distributed? Will it be distributed under a copyright or intellectual property license or terms of use and have any other parties imposed such restrictions on the dataset or any instances in it? Do export controls and other regulatory restrictions apply to the data set or in instances in it? And any further comments regarding distribution of the dataset.
    \item Maintenance: Who will support, host and maintain the dataset and how can this person or these people be contacted? Is there an erratum? Will the data set be updated in the future? If so, how often, by whom and how will these updates be communicated? If the data set includes data relating to people, are there any limits to the amount of time that the data can be retained for? Will older versions of the data set be supported, hosted and maintained? Are there mechanisms for others to extend, augment, build on or contribute to the data set and will their contributions be validated? And any further comments.
\end{itemize}
Some of the security relevant information regarding a dataset, as discussed in previous sections, is already included in the Datasheets for Datasets standard. There are already questions regarding the access to the dataset outside of the organization it was created for, the amount of sensitive data that is part of the dataset and the preprocessing of the data. Dataset maintenance and support are also already generally adressed, although not specifically regarding dataset security. Therefore only a few questions that are more specifically security-focused need to be added to the standard to achieve high quality security documentation for datasets.

\subsection{Adding Security Questions to Datasheets for Datasets}
We propose adding security specific questions to the Datasheets for Datasets standard in multiple sections. Some of these build on existing questions in the standard, whereas others cover entirely new aspects. These are the questions we propose to add and in which section of the datasheet they should be placed:
\begin{itemize}
    \item Motivation: Adding to the first question asked in the Datasheets standard, an additional question should be: “Does the purpose for which the dataset was created put it at an increased risk of adversarial attacks?”. Examples of this would be data sets used for financial calculations, law enforcement, or other areas with a higher risk of attacks.
    \item Preprocessing/Cleaning/Labeling: Adding to the question of whether and how the data was processed, a security specific question “Were mitigations against adversarial attacks conducted?” should be added. Thus practitioners can state whether training data sanitization, adversarial training and other mitigations that work through processing the data have been applied.
    \item Uses: Focused on security the question “Is there a list of known previous successful attacks on the dataset and any malicious accesses to it?” should be added to the Uses section. This would allow dataset consumers, especially of popular and often used datasets, to get an insight into whether the security of an existing data set has already been compromised. 
    \item Maintenance: Regarding the maintenance section, the question of “Will there be security updates to the dataset?” should be added as a more security-specific addition to the existing question regarding updates. The same further questions as for more general updates to the dataset should also be answered regarding security updates, like how often these updates will occur.  
\end{itemize}

\section{Challenges in Implementing the Proposed Method}
Including the security section in Model Cards and security questions in Datasheets for Datasets proposed here pose their own challenges, while some issues regarding the existing standards also remain. One new challenge is the risk of stating security relevant information in documentation that is accessible to potential attackers, particularly for publicly available and open source models. Having such information accessible to adversaries might aid them in creating adversarial examples and other targeted attacks, increasing security risks as opposed to avoiding them. Another new issue might be that the necessary information to complete the security sections may not be available to the practitioners creating the documentation, possibly leading to incomplete or even false documentation.\par
Also, the issue of practitioners not creating documentation at all or creating incomplete documentation will likely remain, as the proposed method does not address this concern. Interactive and automated approaches for it should be developed to allow for easier adoption and use. Additionally, since the existing standards are already not regularly used by practitioners \cite{Liang7}, this problem will most likely also apply to the proposed expanded documentation standards here. And as documentation needs to continuously be updated to be effective \cite{kruse48}, strategies to incentivize practitioners to do so for the proposed method need to be researched as well. To decrease the amount of effort required from practitioners in creating documentation for their ML-systems, a possible future direction of research would be to evaluate whether both dataset and model documentation, including security information, could be integrated into one artifact. Such an artifact, when also implemented interactively, would likely significantly reduce the amount of effort required for completing ML-system documentation. It would also better represent the realities of ML in practice, as models and datasets are regularly not utilized alone, but as part of more complex ML-systems \cite{Apruzzese20}.

\section{Conclusion and Future Work}
To address the problems of low quality ML-documentation and lacking awareness of ML-security, which reinforce each other, the here presented expanded documentation standards can be utilized. Switching from the existing Model Cards and Datasheets for Datasets standards to the expanded versions discussed here is simple for both researchers and practitioners. While this alone is far from enough to contend with the broader challenges in creating secure ML-systems, expanding documentation is one step in the right direction. But more research and well-defined security best practices for ML are clearly needed, both to address the big security challenges facing ML and to further improve upon the here presented security additions to documentation standards. The idea of introducing security sections into existing documentation standards should also be used for the other mentioned ML-documentation standards as well as standards developed in the future.\par
Further research should also be done into the effectiveness of these additions and the success or failure of their adoption by practitioners. The method proposed here of including security information in ML-documentation needs to be evaluated and iterated on both regarding its implementation in practice and through further theoretical research. One possible direction for such an evaluation may be working with organizations which use and create ML-systems and could make the proposed method mandatory for a subset of their employees. Both their ML-systems not documented with the proposed method and those documented with it could then be attacked with the same types of targeted and untargeted attacks, and the success rate of these attacks measured and contrasted. This would give some insight into whether the proposed method is effective in increasing the security of the ML-systems documented with it, through engaging the practitioners creating documentation in actively thinking about security. Also, the amount of security information included in both types of documentation could be compared, to see whether the proposed method is successful at increasing the level of security detail included in documentation.\par
Our proposed method outlines only the integration of security aspects in general ML-documentation. However, as discussed in chapter III, documentation can aid security best when a combined approach of using and creating both specific security documentation and more general documentation including security aspects is used. As there are no specific security documentation standards for ML yet, this remains open as another recommended direction for future research.\par
Better documentation is only one component of increasing ML-security and raising practitioners' awareness for it. Security for ML-systems needs more attention from researchers and practitioners, which more extensive coverage in universities can encourage. Organizations and companies should prioritize security and documentation to address the growing threats against ML-systems. They should encourage the use of existing frameworks and standards or make their use mandatory, so that the existing security best practices and strategies are utilized to their full extent.

\end{document}